\documentclass[conference]{IEEEtran}
\IEEEoverridecommandlockouts

\usepackage{cite}
\usepackage{amsmath,amssymb,amsfonts}
\usepackage{graphicx}
\usepackage{booktabs}
\usepackage{url}
\usepackage{tikz}
\usetikzlibrary{arrows.meta,positioning,shapes.geometric,fit,calc}
\usepackage{listings}
\usepackage{xcolor}
\usepackage{algorithm}
\usepackage{algorithmic}

\lstdefinelanguage{Solidity}{
  keywords={contract,function,external,internal,public,view,returns,struct,enum,mapping,address,uint256,uint64,uint8,bytes32,bytes,bool,if,revert,emit,event,error,calldata,storage,memory,require,constructor},
  sensitive=true,
  morecomment=[l]{//},
  morestring=[b]"
}
\newcommand{\sys}{Attestream}

\begin{document}

\title{\sys{}: Usage-Aware Intermittent Data Distribution with\\
Verifiable Lifecycle Provenance for Machine-Learning Data Streams}

\author{
\IEEEauthorblockN{Kentaro Oda}
\IEEEauthorblockA{\textit{Kagoshima University}\\
Kagoshima, Japan\\
odaken@cc.kagoshima-u.ac.jp}
}

\maketitle

\begin{abstract}
Providers of continuously produced, commercially valuable data---sensor
streams, telemetry, transaction logs, and other feeds sold as
machine-learning training material---face two coupled problems: they cannot
observe whether delivered data is actually used by consumers, and data that
keeps flowing to inactive consumers enlarges the leakage surface without
producing value. We present \sys{}, a blockchain-based
distribution architecture for \emph{intermittently delivered dataset streams}
that couples continued delivery to verifiable usage reporting. Every lifecycle
event---dataset preparation, dual-signed delivery, derivative creation (e.g., a
trained model), and derivative distribution---is appended to an on-chain
registry as a non-repudiable, mutually linked \emph{lifecycle record}. What
the mechanism requires is provable transfer, not tokenization: we define the
record properties abstractly and show that plain contract storage, ERC-721
tokens, and off-chain receipts anchored on-chain are interchangeable
representations of the same protocol. A smart-contract \emph{usage-aware gate} automatically suspends a
consumer's stream when no derivative-creation record is registered for the most
recent delivery within a reporting window, converting provenance registration
from a voluntary courtesy into an economically enforced obligation. A
\emph{modality-pluggable fingerprinting layer} binds any leaked copy to the
dual-signed delivery record of the responsible consumer; we instantiate it
for tabular/geospatial records (attribute perturbation), images
(spread-spectrum watermarking), and documents (randomized fingerprinting
codes). We formalize the gating rule, implement the complete registry as a
Solidity contract with EIP-712 dual signatures, and evaluate it on an EVM
testbed. A full lifecycle round costs $657\mathrm{k}$ gas with plain
records ($\approx\$0.13$ on contemporary rollups; tokenizing the records as
ERC-721 adds ${\sim}30\mathrm{k}$ gas per record, 18\% per round), and the
gate itself adds no dedicated transactions, as it is evaluated lazily
inside delivery registration. Over a
50-consumer pool, leak attribution reaches 100\% accuracy from 40 leaked
table rows under moderate noise, survives JPEG re-compression to quality~30
for images, and tolerates paraphrase rates up to 30\% for documents. The
gating mechanism originates in Japanese patent
JP\,7894573\,B2; this paper contributes its first open formalization,
realization, and quantitative evaluation, together with the
representation analysis and the cross-modality attribution layer.
\end{abstract}

\begin{IEEEkeywords}
blockchain, data provenance, smart contracts, verifiable records, data
marketplaces, machine learning, usage control, data leakage
\end{IEEEkeywords}

\section{Introduction}
\label{sec:intro}

Machine-learning pipelines are increasingly fed by \emph{data streams} rather
than static corpora. An industrial operator sells daily equipment-telemetry
batches to analytics firms that retrain condition-monitoring models as new
data arrives; a mobility provider licenses trip records for demand
forecasting; an environmental-sensing operator supplies each day's
measurements to forecasting services. Across these
settings the asset is not one dataset but an open-ended sequence of small,
timely deliveries, each of which loses value quickly if unused, and each of
which is typically consumed by \emph{incremental retraining} of the buyer's
models.

Existing infrastructure serves this economy poorly on two fronts.
First, \emph{the provider is blind after delivery}. Blockchain provenance
systems \cite{provchain2017,medrec2016,medshare2017,neisse2017} and data
marketplaces \cite{idmob2018,sdte2020,sdpp2019} create tamper-evident records
of \emph{access} or \emph{sale}, but nothing tells the provider whether the
consumer actually exploited the data---information the provider needs to
curate future collection, to price fairly, and to justify continued
investment in sensing. Second, \emph{unused data is pure risk}. Every
delivered copy that sits idle at a consumer is an additional locus from which
the dataset can leak, while generating no corresponding value. A rational
provider would like to stop supplying consumers who no longer use the
stream---but has no signal on which to act.

Our key observation is that these two problems solve each other when usage
reporting is made a \emph{condition of continued supply}. If the consumer must
register a signed, on-chain record of each derivative work (e.g., an updated
model) created from the latest delivery in order to receive the next one, then
(i)~the provider continuously learns whether and how its data is used, and
(ii)~streams to inactive consumers wither automatically, shrinking the leakage
surface precisely where data produces no value. Because each delivery is
intermittent and individually small, the sanction---suspension of the next
delivery---is proportionate, immediate, and enforceable by a smart contract
without litigation.

This paper makes that observation concrete. We present \sys{}, a
blockchain-based distribution architecture with the following contributions:

\begin{itemize}
\item \textbf{Lifecycle provenance as linked verifiable records.} We model
the four events of the data lifecycle---preparation, delivery, derivative
creation, and derivative distribution---as dual-signed records whose fields
cross-reference one another, yielding three on-chain graphs: the temporal
chain of deliveries in a stream, the data-to-derivative usage graph, and the
parent--child lineage of incrementally retrained models. The required
properties (integrity, non-repudiation, linkability) are independent of
representation; tokenization \`a la ERC-721 \cite{erc721} is an optional
add-on whose cost we quantify (Section~\ref{sec:design}).
\item \textbf{A usage-aware delivery gate.} We formalize the rule that
suspends a stream when the latest delivery is not acknowledged by a
derivative-creation record within a reporting window, and show how it can be
evaluated \emph{lazily} inside the next delivery transaction, so that
monitoring adds no dedicated transactions and no off-chain trusted monitor
(Section~\ref{sec:gate}).
\item \textbf{Dual-signed, fingerprint-bound deliveries across
modalities.} Delivery records embed both the provider's transaction
signature and the consumer's EIP-712 \cite{eip712} consent signature,
making them non-repudiable by either party. A modality-pluggable
fingerprinting layer binds leaked copies to a specific dual-signed record;
going beyond the numeric-perturbation embodiment of the prior patent
\cite{patent2023},
we instantiate it for tabular/geospatial records, images (spread-spectrum
watermarking \cite{cox1997}), and documents (randomized fingerprinting
codes \cite{boneh1998,tardos2008}) behind one attribution interface
(Section~\ref{sec:fingerprint}).
\item \textbf{Implementation and evaluation.} We implement the complete
registry as a single Solidity contract and measure per-operation gas,
end-to-end round latency, lineage-query scaling, and attribution accuracy
of all three fingerprint instantiations under partial leakage, adversarial
noise, compression, paraphrasing, and two-party collusion
(Sections~\ref{sec:impl}--\ref{sec:eval}).
\end{itemize}

The usage-gated distribution mechanism that \sys{} realizes was first
proposed in Japanese patent JP\,7894573\,B2 \cite{patent2023} (published
as application JP\,2023-43186\,A in 2023 and granted in 2026), held by
Kagoshima University, Ocean Solution
Technology, and Lily; the present author is one of its inventors. That
document specifies the system at the level of functional units and
embodiments, but contains no formal gating rule, no analysis of the record
representation, no concrete realization, and no measurements. Building on
it as prior work, this paper contributes the formalization, the
representation-agnostic record abstraction and its cost ablation, the lazy
gate realization, the modality-pluggable fingerprinting layer, and the
quantitative evaluation---none of which appear in the patent literature.

\section{Related Work}
\label{sec:related}

\subsection{Blockchain Data Provenance and Sharing}
ProvChain \cite{provchain2017} anchors cloud file-operation logs in a
blockchain; MedRec \cite{medrec2016} manages medical-record access permissions
via Ethereum contracts; MeDShare \cite{medshare2017} monitors inter-provider
medical data flows and can revoke access on detected policy violations; Neisse
et al.\ \cite{neisse2017} model data-usage accountability contracts for GDPR
compliance. These systems log or police \emph{access to primary data}. \sys{}
extends provenance across \emph{derivative artifacts} and inverts the
enforcement direction: rather than sanctioning detected misuse, it sanctions
the \emph{absence} of affirmative usage reports, which no prior provenance
system enforces.

\subsection{Data Marketplaces and Trading}
IDMoB \cite{idmob2018} matches IoT vendors and buyers on-chain; SDTE
\cite{sdte2020} secures the trade itself with trusted hardware; SDPP
\cite{sdpp2019} interleaves micropayments with streaming records. All treat
the sale as the terminal event. \sys{} instead treats delivery as the
\emph{beginning} of an obligation cycle---report usage or lose supply---and
extends billing hooks to derivative sales, which marketplaces do not track.

\subsection{Tokenized Records}
Surveys of the NFT ecosystem \cite{wangnft2021} and frameworks for deployable
NFT contracts \cite{chirtoaca2020} focus on tokens as tradeable
representations of single assets. \sys{} needs none of that machinery: its
records are \emph{dual-signed lifecycle events} whose value lies in their
links to one another, and verifiability---not transferability---is the
point. Tokenization is retained only as an optional representation for
ecosystems that want wallet visibility or royalty rails, and
Section~\ref{sec:eval} quantifies exactly what that option costs.

\subsection{Model and Dataset Provenance}
Datasheets \cite{datasheets2021} and model cards \cite{modelcards2019} are
voluntary, off-chain documentation without authenticity guarantees. Lo et
al.\ \cite{lo2023} record federated-learning provenance on-chain among
cooperating parties. \sys{} targets the adversarial provider--consumer
setting and makes dataset-to-model lineage a cryptographically signed record
whose registration is compelled by the delivery gate.

\subsection{Access and Usage Control via Smart Contracts}
FairAccess \cite{fairaccess2016}, contract-based IoT access control
\cite{zhang2019}, auditable policy contracts \cite{maesa2019}, and Droplet's
cryptographic stream authorization \cite{droplet2020} all gate on \emph{who}
may access a resource, or penalize detected misbehavior \cite{zhang2019}.
\sys{}'s gate is, to our knowledge, the first to suspend supply on a
consumer's \emph{failure to act}---a continuous-obligation semantics closer
to usage control (UCON) than to access control, realized natively in the
delivery path.

\subsection{Leak Tracing and Accountability}
Media watermarking \cite{cox1997}, document marking \cite{brassil1995},
database watermarking and fingerprinting \cite{agrawal2002,li2005}, GIS
coordinate watermarking \cite{abubahia2017}, collusion-secure
fingerprinting codes \cite{boneh1998,tardos2008}, and radioactive data
\cite{radioactive2020} identify a guilty recipient after a leak;
probabilistic leakage attribution \cite{papadimitriou2011} does so without
marks; LUCE \cite{luce2019} monitors license compliance on-chain. \sys{} does not propose new marks; its contribution is to place existing
embedders behind one attribution interface and to combine ex-ante reduction
of the leak surface (unused streams stop flowing) with ex-post attribution
whose evidence---the dual-signed delivery record of the fingerprinted
copy---is already on-chain. In application domains such as
environmental sensing and supply chains, blockchain has been applied to
product traceability \cite{probst2020,alsharabi2024}, not to sensor-data
monetization with derivative tracking.

\section{System Model and Design Goals}
\label{sec:model}

\subsection{Actors and Assets}
A \emph{provider} $P$ prepares, at irregular intervals (every one to a few
days), a fresh dataset $D_j$---the $j$-th element of an open-ended stream
$\mathcal{D} = D_1, D_2, \ldots$. Each registered \emph{consumer} $C_i$
receives its own fingerprinted variant $D_j^{(i)}$ and may use it, possibly
together with its previously created derivative (e.g., yesterday's model), to
produce a new \emph{derivative} $M$. Consumers may further distribute
derivatives to third-party \emph{recipients}. A \emph{history-management
ledger} $\mathcal{L}$---a blockchain hosting the \sys{} contract---stores all
lifecycle records. Payloads themselves are exchanged off-chain (direct
download or encrypted object storage); the ledger stores only hashes,
identities, timestamps, links, and signatures.

\subsection{Threat Model}
Provider and consumers are mutually distrusting rational parties. A consumer
may (i) deny having received a delivery, (ii) use data without reporting,
(iii) leak its copy, or (iv) register fabricated usage reports. A provider may
(v) deny having delivered, or (vi) falsely accuse a consumer of leaking. The
ledger is assumed to provide integrity and availability (standard blockchain
assumptions \cite{wood2014}); signing keys are not compromised. Verifying
\emph{semantic} truth of a usage report (that a registered hash really is a
model trained on the data) is out of scope of the on-chain mechanism and is
discussed in Section~\ref{sec:discussion}.

\subsection{Design Goals}
\begin{enumerate}
\item[\textbf{G1}] \emph{Usage visibility:} the provider can determine, per
consumer and per delivery, whether the data was used to create a derivative.
\item[\textbf{G2}] \emph{Automatic supply discipline:} streams to consumers
who stop reporting usage are suspended without provider intervention or a
trusted monitor.
\item[\textbf{G3}] \emph{Non-repudiation:} neither party can later deny a
completed delivery; consumers cannot deny authorship of registered
derivatives.
\item[\textbf{G4}] \emph{Lineage:} every derivative is linked to the
deliveries it consumed and to its parent derivative, transitively.
\item[\textbf{G5}] \emph{Leak attribution:} a leaked payload identifies the
responsible consumer, with on-chain evidence.
\end{enumerate}

\section{The \sys{} Design}
\label{sec:design}

\begin{figure}[t]
\centering
\begin{tikzpicture}[
  font=\scriptsize,
  box/.style={draw, rounded corners=1pt, minimum height=6mm, inner sep=2pt, align=center},
  rec/.style={box, fill=gray!12, minimum width=17mm},
  actor/.style={box, thick},
  lnk/.style={-{Stealth[length=1.8mm]}, thin},
  dsh/.style={-{Stealth[length=1.8mm]}, densely dashed, thin}
]
\node[actor] (prov) at (0.6,2.2) {Provider $P$};
\node[actor] (cons) at (4.2,2.2) {Consumer $C_i$};
\node[actor] (thrd) at (7.4,2.2) {Recipient $R$};
\node[rec] (tr1) at (0.6,0)  {TR1 Prep.\\$\rho_j$};
\node[rec] (tr2) at (2.9,0)  {TR2 Delivery\\$\delta_j^{(i)}$ dual-sig.};
\node[rec] (tr3) at (5.2,0)  {TR3 Deriv.\\creation $\mu_k$};
\node[rec] (tr4) at (7.5,0)  {TR4 Deriv.\\distribution};

\node[draw, densely dotted, inner sep=4pt, fit=(tr1)(tr2)(tr3)(tr4),
      label={[font=\scriptsize, yshift=-6.5mm]below:{Blockchain ledger $\mathcal{L}$ (\sys{} contract)}}] {};

\draw[dsh] (prov) -- node[above]{payload $D_j^{(i)}$} (cons);
\draw[dsh] (cons) -- node[above]{model $M_k$} (thrd);

\draw[lnk] (prov.south) -- node[left]{reg.} (tr1.north);
\draw[lnk] (prov.south east) -- node[pos=.45, below, sloped]{reg.} (tr2.north west);
\draw[lnk] (cons.south west) -- node[pos=.45, above, sloped]{consent} (tr2.north east);
\draw[lnk] (cons.south east) -- node[pos=.45, below, sloped]{reg.} (tr3.north west);
\draw[lnk] (cons.east) to[out=-25,in=115]
  node[pos=.72, above, sloped]{reg.} (tr4.north west);
\draw[lnk] (thrd.south) -- node[pos=.6, right]{consent} (tr4.north east);

\draw[lnk] (tr2.south) to[out=-90,in=-90, looseness=0.9]
  node[below, pos=.5]{used-prep} (tr1.south);
\draw[lnk] (tr3.south) to[out=-90,in=-90, looseness=0.9]
  node[below, pos=.5]{used-delivery} (tr2.south);
\draw[lnk] (tr4.south) to[out=-90,in=-90, looseness=0.9]
  node[below, pos=.5]{derivative} (tr3.south);
\draw[lnk] (tr2.west) to[out=150,in=90, looseness=3]
  node[above, pos=.5, xshift=1mm]{prev} (tr2.north);
\draw[lnk] (tr3.east) to[out=30,in=90, looseness=3]
  node[above, pos=.5, xshift=-1mm]{parent} (tr3.north);
\end{tikzpicture}
\caption{\sys{} overview. Payloads move off-chain (dashed); every lifecycle
event appends a linked, dual-signed record on-chain. The delivery gate (Alg.~\ref{alg:gate})
is evaluated inside TR2 registration.}
\label{fig:arch}
\end{figure}
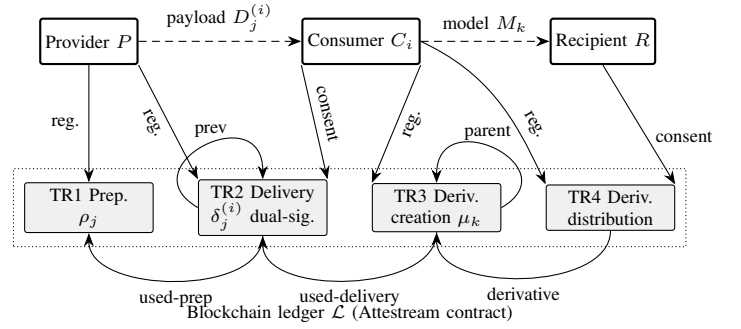

\subsection{Lifecycle Records as a Linked Provenance Graph}
\label{sec:records}

\sys{} registers four record types, mirroring transactions TR1--TR4 of the
patent \cite{patent2023} (Fig.~\ref{fig:arch}):

\subsubsection{Preparation (TR1)} When $P$ finishes assembling $D_j$, it registers
$\rho_j = (\mathrm{hash}(D_j), P, t)$. This timestamps authorship of the raw
dataset before any delivery, analogous to a notarized deposit.

\subsubsection{Delivery (TR2)} After $C_i$ downloads its fingerprinted variant
$D_j^{(i)}$, $P$ registers
$\delta_j^{(i)} = (P, C_i, \mathrm{hash}(D_j^{(i)}), \rho_j,
\delta_{j-1}^{(i)}, t, \sigma_{C_i})$,
where $\sigma_{C_i}$ is the consumer's EIP-712 signature over
$(P, C_i, \mathrm{hash}(D_j^{(i)}), \rho_j, n)$ with a per-consumer nonce $n$
preventing replay. The contract verifies $\sigma_{C_i}$ on-chain; the
transaction itself carries $P$'s signature, so the confirmed record is signed
by \emph{both} parties (G3). The $\delta_{j-1}^{(i)}$ back-pointer chains all
deliveries of one stream (the patent's ``previous-data identification'').

\subsubsection{Derivative creation (TR3)} When $C_i$ trains a model $M_k$ from
$D_j^{(i)}$ (optionally refining a previous model $M_{k-1}$), it registers
$\mu_k = (C_i, \mathrm{hash}(M_k), \delta_j^{(i)}, \mu_{k-1}, t)$.
The \emph{used-delivery} link realizes G1 and G4; the \emph{parent} link
records incremental-training lineage, so the full ancestry of any model
version is walkable on-chain.

\subsubsection{Derivative distribution (TR4)} When $C_i$ transfers $M_k$ to a
recipient $R$, it registers a record carrying both $C_i$'s and $R$'s
signatures, extending non-repudiation one hop down the value chain and
enabling royalty-style billing on derivative sales.

Only the record \emph{owner} roles can register: TR1/TR2 by the provider (TR2
additionally requiring the consumer's consent signature), TR3 by the consumer
named in the referenced delivery, TR4 by the creator of the referenced
derivative. All references are validated for type and ownership at registration time,
so the on-chain graph is well-formed by construction.

\subsection{Record Representation}
\label{sec:representation}

The protocol requires of a record store only four properties:
\emph{integrity} (records cannot be altered once confirmed),
\emph{non-repudiation} (each record binds the signatures of its parties),
\emph{linkability} (records reference one another by stable identifiers),
and \emph{queryable availability} (the gate and lineage walks can read
them). It does \emph{not} require token semantics---no ownership transfer,
no wallet balance, no marketplace interface. The patent itself spans this
spectrum: its first embodiment stores history records in an audited
third-party database, and only its fourth realizes them as NFTs. We
therefore treat the representation as a deployment choice:

\subsubsection{Plain contract records (default)} Each record is a struct in
contract storage plus an event; identifiers are sequence numbers. This is
the minimal on-chain realization of the four properties and the cheapest
(Section~\ref{sec:eval-repr}).

\subsubsection{ERC-721 tokenization (optional)} The same structs are
additionally minted as ERC-721 tokens \cite{erc721}. This buys wallet and
explorer visibility, standardized custody, and composability with
royalty/marketplace rails---relevant if derivative records are themselves
traded---at a measurable gas premium.

\subsubsection{Anchored receipts / audited database} Where no chain is
desired, dual-signed receipts can live in an audited database (the patent's
first embodiment), optionally anchored by periodic hash commitments to a
public chain; integrity then rests on the auditor plus the anchors, and
the gate runs in the distribution server rather than a contract.

The gate logic, signature scheme, link structure, and fingerprinting layer
are identical across representations; our prototype implements the first
two on the same code path and we quantify the difference in
Section~\ref{sec:eval-repr}.

\subsection{The Usage-Aware Delivery Gate}
\label{sec:gate}

Let $w$ be the \emph{reporting window} agreed at stream registration
(shorter than the delivery interval; e.g., $w = 1$ day for a daily stream).
Let $t_j$ be the confirmation time of delivery $\delta_j^{(i)}$ and
$a_j \in \{\bot\} \cup \mathbb{R}$ the time at which some TR3 record first
referenced $\delta_j^{(i)}$ ($\bot$ if none). Define
\begin{equation}
\mathrm{ack}(j) \;\equiv\; a_j \neq \bot \;\wedge\; a_j \le t_j + w .
\label{eq:ack}
\end{equation}
The gate admits delivery $j{+}1$ at time $t$ iff the stream is registered,
not administratively suspended, and
\begin{equation}
j = 0 \;\vee\; \mathrm{ack}(j) \;\vee\; t \le t_j + w .
\label{eq:gate}
\end{equation}
The last disjunct lets an early next delivery proceed while the window for
the previous one is still open; once the window has lapsed without
acknowledgment, every subsequent delivery attempt fails and the stream is
marked suspended until the provider explicitly reinstates it. A \emph{late}
TR3 record (registered after $t_j + w$) does not reopen the gate---the
consumer demonstrated non-compliance, and reinstatement is a provider
decision, mirroring the patent's provider-controlled delivery-permission
flag.

\begin{algorithm}[t]
\caption{Lazy gate evaluation inside \texttt{recordDelivery}}
\label{alg:gate}
\begin{algorithmic}[1]
\REQUIRE stream state $s = (\delta_{\mathrm{last}}, t_{\mathrm{last}},
\mathit{ack}, a_{\mathrm{last}}, \mathit{susp})$, window $w$, now $t$
\IF{$\neg s.\mathrm{registered} \lor s.\mathit{susp}$} \RETURN revert
\ENDIF
\IF{$\delta_{\mathrm{last}} \neq \bot$}
  \STATE $\mathit{timely} \gets \mathit{ack} \wedge (a_{\mathrm{last}} \le t_{\mathrm{last}} + w)$
  \IF{$\neg\mathit{timely} \wedge t > t_{\mathrm{last}} + w$}
    \STATE $s.\mathit{susp} \gets \mathrm{true}$; \textbf{emit} Suspended; \RETURN revert
  \ELSIF{$\neg\mathit{timely}$} \RETURN revert \COMMENT{window still open}
  \ENDIF
\ENDIF
\STATE verify consumer EIP-712 consent; append $\delta$; update $s$
\end{algorithmic}
\end{algorithm}

A naive realization would run an off-chain monitor (the patent's
``monitoring unit'') that polls the ledger at $t_j + w$ and writes a
permission flag---one extra transaction per consumer per round, plus a
trusted scheduler. Our key implementation insight is that the gate predicate
\eqref{eq:gate} depends only on \emph{stored state and the current block
timestamp}, so it can be evaluated lazily inside the next
\texttt{recordDelivery} call (Alg.~\ref{alg:gate}): suspension takes effect
exactly when it matters---at the moment the provider attempts the next
delivery---at zero additional transaction cost, and \texttt{deliveryAllowed}
remains available as a free \texttt{view} call for off-chain pre-checks
(G2). Block timestamps are miner-influenceable only by seconds, negligible
against windows of hours or days.

\subsection{Modality-Pluggable Fingerprinting and Leak Attribution}
\label{sec:fingerprint}

The patent's second embodiment derives each consumer's variant by
perturbing numeric attributes of tabular records. We generalize this into
a pluggable \emph{fingerprint embedder} abstraction, so that one
attribution pipeline covers the payload types that real data streams
actually carry---tables, images, and documents. An embedder is a keyed
function $F$ mapping the master $D_j$ and a consumer key $k_i$ to the
variant $D_j^{(i)} = F(D_j, k_i)$ satisfying three properties:
\emph{fidelity} ($D_j^{(i)}$ retains the utility of $D_j$),
\emph{robustness} (the mark survives the transformations a leaker
plausibly applies), and \emph{distinguishability} (variants of different
consumers are reliably separable even from partial content). The on-chain
side is modality-agnostic: the contract only ever sees
$\mathrm{hash}(D_j^{(i)})$ and maintains the index
$\mathrm{hash}(D_j^{(i)}) \mapsto \delta_j^{(i)}$. We instantiate $F$ for
three modalities:

\subsubsection{Tabular and geospatial records (the patent's embodiment)}
Zero-mean Gaussian perturbation of error-tolerant numeric attributes
\cite{agrawal2002,li2005}, seeded by $k_i$. Our reference instantiation
perturbs the latitude/longitude of geotagged records by
$\mathcal{N}(0, \sigma^2)$ with $\sigma \approx 0.0005^\circ$
($\approx$55\,m), well within typical sensor tolerance \cite{abubahia2017};
measurements, timestamps, or least-significant digits can carry the mark
instead. Attribution matches leaked rows against each consumer's expected
variant by mean squared error.

\subsubsection{Images}
A Cox-style multiplicative spread-spectrum watermark \cite{cox1997}: the
$M$ perceptually most significant mid-frequency DCT coefficients $v$ of the
master are replaced by $v(1 + \alpha w_i)$, where $w_i \sim \mathcal{N}(0,1)^M$
is pseudorandomly derived from $k_i$ ($\alpha = 0.1$, $M = 4096$ in our
prototype). Attribution extracts the relative coefficient deviations of a
leaked image and correlates them against each candidate pattern; spread-%
spectrum embedding makes the mark robust to compression, filtering, and
requantization while remaining imperceptible.

\subsubsection{Documents}
Text does not tolerate numeric perturbation, but exposes
\emph{meaning-preserving binary choice sites}: synonym pairs, equivalent
phrasings, punctuation and formatting variants \cite{brassil1995}. Each
consumer's variant realizes the $S$ sites of the master according to a
pseudorandom codeword $c_i \in \{0,1\}^S$ derived from $k_i$---a randomized
fingerprinting code in the sense of Boneh--Shaw and Tardos
\cite{boneh1998,tardos2008}. Attribution over a leaked excerpt scores the
agreement of the visible sites with each candidate codeword; the code-based
formulation additionally yields resistance to small coalitions of colluding
consumers, which per-record perturbation alone does not provide.

In all three cases the provider recomputes (or, for partial leaks,
statistically matches, Section~\ref{sec:eval-fp}) the fingerprint of a
leaked file and presents the dual-signed delivery record as evidence that
\emph{this consumer received exactly this copy} (G5). Because the consumer
co-signed $\mathrm{hash}(D_j^{(i)})$ at delivery time, it cannot claim the
fingerprint was planted after the fact---an evidentiary link that
standalone watermarking schemes lack.

\subsection{Billing Hooks}
Delivery and derivative-distribution records give the provider a
non-repudiable basis for per-delivery charging and for revenue sharing on
derivative sales (the patent's third embodiment). Settlement can be off-chain
against on-chain evidence, or native (escrowed payment released by the TR2
registration); we implement records only and discuss settlement as
orthogonal.

\section{Implementation}
\label{sec:impl}

We implemented the registry as Solidity 0.8.24 contracts on OpenZeppelin's
EIP-712 (and, for the tokenized variant, ERC-721) libraries, compiled with
the optimizer at 200 runs for the Cancun EVM. Two variants share one code
path (${\sim}320$ lines each): the default stores each record as a struct
in contract storage plus an event, and the tokenized variant additionally
mints each record as an ERC-721 token
(Section~\ref{sec:representation}). In both, a \texttt{Record} struct
stores type, actor addresses, payload hash, timestamp, and the three link
fields (\texttt{prev}, \texttt{used}, \texttt{parent}). Stream state is a
\texttt{(provider, consumer)}-keyed mapping holding the window, last
delivery id and time, and acknowledgment state; TR3 registration updates the
acknowledgment in $O(1)$. Consent signatures use EIP-712 typed data with
per-consumer nonces. The listing below shows the gate on the hot path:

\begin{lstlisting}[language=Solidity]
bool timely = s.lastAck
  && s.ackTime <= s.lastTime + s.window;
bool open = block.timestamp
  <= s.lastTime + s.window;
if (!timely && !open) {
  s.suspended = true;
  emit StreamSuspended(provider, consumer,
                       s.lastDeliveryId);
  revert ReportingWindowElapsed(
    s.lastDeliveryId);
}
if (!timely) revert
  ReportingWindowElapsed(s.lastDeliveryId);
\end{lstlisting}

The prototype comprises both contract variants, a seven-case property test
suite (lifecycle round-trip, gating on silence, late-acknowledgment
handling, multi-round streams, lineage walking, forged-consent rejection,
and unregistered-consumer rejection) executed against each variant (14
runs, all passing), and the evaluation harness. Source code is available
from the author.

\section{Evaluation}
\label{sec:eval}

We ask: (\textbf{Q1}) What does lifecycle registration cost on-chain, and is
it viable on public infrastructure? (\textbf{Q2}) How does the lazy gate
affect cost and latency? (\textbf{Q3}) How reliably do the three
fingerprint instantiations attribute partial, degraded leaks? On-chain
experiments ran on a local Hardhat/EDR EVM node (Cancun) with 100 full
lifecycle rounds; gas figures are EVM-deterministic and thus identical on
any EVM chain, while latency figures are testbed-local upper bounds on
throughput, in practice dominated by chain block times. All fingerprint
experiments use a pool of 50 registered consumers (chance level 2\%).

\subsection{On-Chain Cost (Q1, Q2)}

\begin{figure}[t]
\centering
\includegraphics[width=\columnwidth]{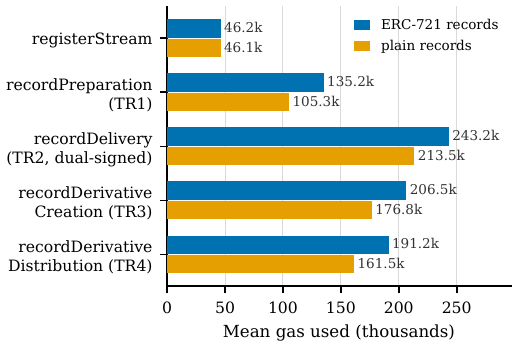}
\caption{Mean gas per operation over 100 lifecycle rounds, for the plain
and ERC-721 record representations. Round-to-round variation is below
0.1\% except for first-round cold storage-slot initialization, which
raises the maxima by up to 13\%.}
\label{fig:gas}
\end{figure}

\begin{table}[t]
\caption{Gas per operation for both record representations (means;
$n=100$ rounds, $n=20$ for \texttt{registerStream}), and projected cost of
the plain variant (Ethereum L1 at 5\,gwei, rollup at 0.05\,gwei,
ETH = \$4{,}000).}
\label{tab:gas}
\centering
\begin{tabular}{@{}lrrrr@{}}
\toprule
Operation & Plain & ERC-721 & L1 (USD) & Rollup \\
\midrule
Contract deployment & 1{,}521{,}789 & 2{,}318{,}786 & 30.44 & 0.30 \\
\texttt{registerStream} & 46{,}145 & 46{,}235 & 0.92 & 0.009 \\
TR1 \texttt{recordPreparation} & 105{,}297 & 135{,}163 & 2.11 & 0.021 \\
TR2 \texttt{recordDelivery} & 213{,}548 & 243{,}244 & 4.27 & 0.043 \\
TR3 \texttt{recordDerivCreation} & 176{,}768 & 206{,}507 & 3.54 & 0.035 \\
TR4 \texttt{recordDerivDistrib} & 161{,}473 & 191{,}206 & 3.23 & 0.032 \\
\midrule
Full round (TR1--TR4) & 657{,}086 & 776{,}120 & 13.14 & 0.131 \\
\bottomrule
\end{tabular}
\end{table}

Table~\ref{tab:gas} and Fig.~\ref{fig:gas} summarize per-operation gas.
TR2 is the most expensive operation (214\,k gas in the plain
representation): it verifies an EIP-712 signature, checks the gate, appends
the record, and writes the delivery index and stream state. A complete
daily round---preparation, delivery, derivative report, and one derivative
distribution---costs 657\,k gas, i.e., $\approx\$13$ on Ethereum L1 under
the stated assumptions but only $\approx\$0.13$ on contemporary rollups and
marginal cost on a consortium chain, comfortably below the value of a
commercial daily data delivery. Costs scale linearly in consumers and
rounds; no operation touches unbounded state.

The lazy gate (Alg.~\ref{alg:gate}) adds only a few storage reads and one
comparison to TR2---within round-to-round noise---and, crucially,
\emph{eliminates} the per-consumer-per-round monitoring transaction of the
naive design, so the monitoring mechanism itself is gas-free
(Q2). Suspension consumes gas only in the exceptional path (one storage
write and event inside the reverting call's gas).

\subsection{Record-Representation Ablation}
\label{sec:eval-repr}

Because the protocol needs provable transfer rather than tokens
(Section~\ref{sec:representation}), we measured both representations on
the same code path. Dropping ERC-721 minting saves a near-constant
${\sim}30$\,k gas per record---the token's ownership and balance
writes plus the transfer event---which amounts to 12--22\% per operation,
15.3\% (119\,k gas) per full round, and 34\% of deployment cost
(Table~\ref{tab:gas}). The gate, signatures, links, and queries are
unaffected. Tokenization is thus purely a feature decision: it is worth
its premium only where wallet visibility, standardized custody, or
royalty rails over derivative records are actually consumed by the
ecosystem, and the numbers above price that decision.

\subsection{Latency and Query Scaling}

\begin{figure}[t]
\centering
\includegraphics[width=\columnwidth]{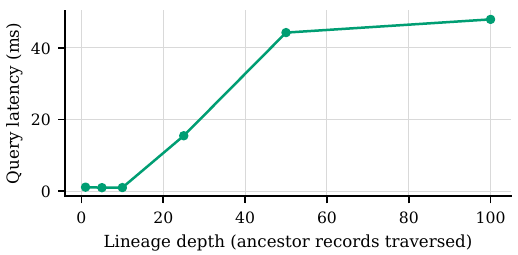}
\caption{Lineage-query latency vs.\ ancestry depth (local node, view call).}
\label{fig:lineage}
\end{figure}

A full four-transaction round completes in 13.1\,ms mean (p95: 18.6\,ms;
plain records; the tokenized variant adds ${\sim}3$\,ms) on
the local node, i.e., the contract logic sustains ${>}75$ rounds/s per
process; real deployments are bounded by block production, not by \sys{}
logic. Walking a model's ancestry via \texttt{lineageOf} (a \texttt{view}
call) returns 100 generations in under 50\,ms (Fig.~\ref{fig:lineage});
leak-attribution lookup by payload hash is a single mapping read
(under 1\,ms including RPC overhead).

\subsection{Fingerprint Attribution Across Modalities (Q3)}
\label{sec:eval-fp}

\subsubsection{Tabular/geospatial records}
\begin{figure}[t]
\centering
\includegraphics[width=\columnwidth]{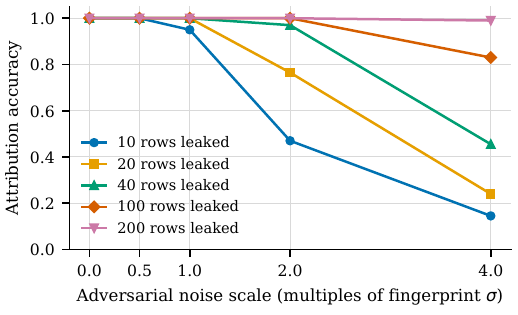}
\caption{Tabular records: attribution accuracy vs.\ adversarial noise, for
varying numbers of leaked rows (200 trials per point).}
\label{fig:fingerprint}
\end{figure}

We simulate a 2{,}000-row dataset of location-stamped records, each
consumer holding a variant fingerprinted with per-consumer Gaussian
coordinate noise ($\sigma = 0.0005^\circ$). A leaker releases a random
subset of rows and may add its own Gaussian noise of scale $\lambda\sigma$
to degrade the mark; attribution minimizes mean squared coordinate error
over the leaked rows. Fig.~\ref{fig:fingerprint} reports accuracy over 200
trials per configuration.

With no or moderate added noise ($\lambda \le 1$), attribution is essentially
perfect from as few as 10--20 leaked rows. Even an adversary injecting noise
at twice the fingerprint scale ($\lambda = 2$)---already doubling the
positional error of its own product---is identified with 97\% accuracy from
40 rows and 100\% from 100 rows. Only at $\lambda = 4$, where the leaked
data's positional quality is severely degraded, does small-leak attribution
fail, and 200+ leaked rows still yield 99\% accuracy.

\subsubsection{Images}
\begin{figure}[t]
\centering
\includegraphics[width=\columnwidth]{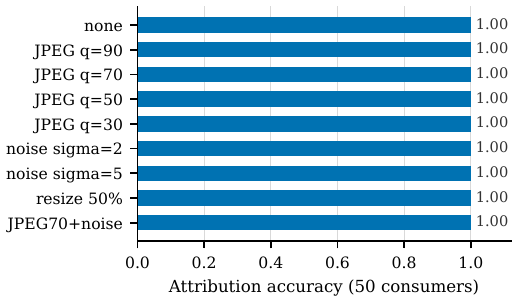}
\caption{Images: attribution accuracy after common removal attacks
(spread-spectrum mark, $\alpha=0.1$, mean PSNR 53\,dB; one trial per
consumer per attack).}
\label{fig:image}
\end{figure}

We embed the spread-spectrum mark ($\alpha = 0.1$, $M = 4096$ DCT
coefficients) into a $512{\times}512$ reference image for each of the 50
consumers; the marked copies are visually indistinguishable from the master
(mean PSNR 53.1\,dB, min 52.3\,dB). Each leaked copy is subjected to an
attack before attribution. Fig.~\ref{fig:image} shows the outcome:
attribution remains at 100\% under every attack tested---JPEG
re-compression down to quality 30, additive pixel noise up to
$\sigma = 5$, a 50\% downscale--upscale cycle, and a combined
JPEG-plus-noise attack---because the 4096-chip spread-spectrum pattern
degrades gracefully while remaining strongly correlated with its key.

\subsubsection{Documents}
\begin{figure}[t]
\centering
\includegraphics[width=\columnwidth]{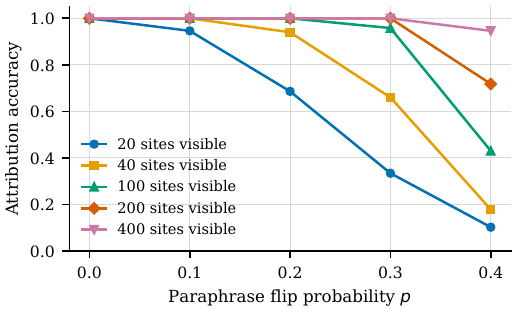}
\caption{Documents: attribution accuracy vs.\ paraphrase flip probability,
for varying numbers of visible choice sites ($S=400$ total, 500 trials per
point).}
\label{fig:text}
\end{figure}

We model a document with $S = 400$ meaning-preserving binary choice sites
and give each consumer a variant realized by its pseudorandom codeword. A
leaker publishes an excerpt exposing a fraction of the sites and applies a
paraphrase pass that flips each visible site with probability $p$;
attribution maximizes codeword agreement over visible sites
(Fig.~\ref{fig:text}, 500 trials per point). With a tenth of the document
visible (40 sites), attribution is 94\% accurate under $p = 0.2$ and
perfect under lighter paraphrasing; with half the document, it withstands
$p = 0.3$ at 100\%. Under a two-consumer collusion in which the coalition
splices its two variants site-by-site, the top-scoring codeword belongs to
a colluder in 96.8\% of trials with 40 visible sites and 100\% from 100
sites---the behavior expected of randomized fingerprinting codes
\cite{boneh1998,tardos2008}, whose collusion bounds sharpen further with
dedicated code constructions.

Across all three modalities, attribution accuracy grows with exactly what
makes a leak harmful---its volume and fidelity---and every identification
is backed by the consumer's own delivery co-signature.

\section{Discussion}
\label{sec:discussion}

\subsection{Security Properties}
\emph{Non-repudiation (G3).} A confirmed TR2 record carries the provider's
transaction signature and the consumer's EIP-712 consent over the payload
hash and preparation reference; nonces prevent replay. Neither party can
later deny the delivery, and disputes reduce to signature verification
against the ledger.
\emph{False accusation resistance.} The provider cannot frame a consumer: a
leaked file attributes only if it matches a fingerprint whose hash the
consumer itself co-signed.
\emph{Gate manipulation.} The consumer cannot forge acknowledgments (TR3
requires its own signature referencing a delivery addressed to it), and the
provider cannot fabricate a missed window, since delivery and
acknowledgment times are consensus timestamps.

\subsection{Limitations}
\emph{Semantic truth of reports.} The contract verifies \emph{that} a
consumer registered a derivative hash, not that the hash denotes a genuine
model trained on the delivered data. A consumer determined to keep the
stream flowing while idling can register junk hashes at a cost of
$\approx 177$\,k gas per round. The gate is therefore an \emph{incentive}
mechanism---it makes silent free-riding impossible and dishonest reporting
attributable and auditable (registered hashes are commitments the provider
may spot-check contractually)---rather than a proof of training.
Cryptographic proofs of training (e.g., zkML) could close this gap and slot
naturally into the TR3 interface.
\emph{Fingerprint robustness.} The three embedders withstand the
subsetting, noise, compression, and paraphrasing attacks evaluated in
Section~\ref{sec:eval-fp}, but each has known stronger adversaries:
aggressive coarsening or aggregation for numeric marks, desynchronizing
geometric transforms (rotation, cropping) for the global-DCT image mark,
and full semantic rewriting for document codes. Hardened embedders from
the respective literatures \cite{li2005,cox1997,tardos2008,radioactive2020}
slot into the same interface without touching the on-chain layer, and
larger collusion coalitions call for dedicated code constructions
\cite{tardos2008}.
\emph{Privacy.} Records expose pseudonymous activity patterns
(who receives, who trains, how often). Deployments over public chains can
blind identities with per-stream addresses; payload content is never
on-chain.
\emph{Collusion.} A consumer can outsource training or launder data through
a colluding recipient; TR4 records extend accountability one hop, but
transitive enforcement beyond direct counterparties remains open.

\subsection{Deployment Considerations}
For a consortium of data providers and analytics firms---the setting that
motivated this design---a permissioned EVM chain gives
negligible marginal cost and full control over validator membership, while a
public rollup provides stronger neutrality at $\approx\$0.13$ per round.
The contract is chain-agnostic EVM bytecode; the reporting window $w$, the
delivery cadence, and the fingerprint scale are per-stream parameters.

\section{Conclusion}
\label{sec:conclusion}

\sys{} turns the provider's blindness after data delivery into an enforced
feedback loop: lifecycle events become dual-signed, mutually linked
verifiable records,
and continued supply is conditioned---by contract logic alone, at zero added
transaction cost---on timely, attributable usage reporting. Evaluation of
our Solidity prototype shows the complete mechanism is practical today:
${\sim}776$\,k gas per lifecycle round, monitoring without dedicated
transactions, and near-perfect leak attribution across tabular, image, and
document payloads from small, degraded fractions of a fingerprinted
delivery. Future work includes proof-of-training integration for
semantically verified reports, hardened and collusion-secure embedders
behind the same interface, privacy-preserving record encodings, and field
deployment on commercial data streams.

\section*{Acknowledgment}
The usage-gated distribution mechanism realized in this paper originates
in Japanese patent JP\,7894573\,B2 \cite{patent2023},
invented jointly with Yosuke Mizukami (Ocean Solution Technology Inc.) and
Hiroyuki Nozaki (Lily Co., Ltd.); the patent is held by Kagoshima
University, Ocean Solution Technology Inc., and Lily Co., Ltd. The author thanks the
co-inventors for the collaboration that produced the original mechanism.
The formalization, implementation, evaluation, and extensions reported
here are the author's own work.

\bibliographystyle{IEEEtran}
\bibliography{refs}

\end{document}